\documentclass[12pt]{article}
\usepackage{amssymb}
\usepackage{a4}
\evensidemargin \oddsidemargin
\def\1{{\bf 1}}
\def\id{\mbox{id\,}}
\def\ot{\otimes}
\def\uot{\,\underline{\otimes}\,}
\def\ots{\otimes_{\star}}
\def\F{\mbox{$\cal F$}}
\def\bF{\mbox{$\overline{\cal F}$}}

\def \D {{\cal D}}
\def \H {{\cal H}}
\def \Ha{{\sf H}}
\def \W {{\cal W}}
\def \G {{\cal G}}
\def\R{\mbox{$\cal R$\,}}

\def \bV {\overline{V}_+}
\def \tW {\tilde{W}}

\def\A{\mbox{$\cal A$}}
\def\hA{\mbox{$\widehat{\cal A}$}}

\newcommand{\tr}{\hat\triangleright}
\newcommand{\trc}{\triangleright}

\def\b#1{{\mathbb #1}}
\newcommand{\be}{\begin{equation}}
\newcommand{\ee}{\end{equation}}
\newcommand{\bea}{\begin{eqnarray}}
\newcommand{\eea}{\end{eqnarray}}
\newcommand{\ba}{\begin{array}}
\newcommand{\ea}{\end{array}}
\def\sq{\mbox{\rlap{$\sqcap$}$\sqcup$}}
\newenvironment{proof}[1]{\vspace{5pt}\noindent{\bf Proof #1}\hspace{6pt}}%
{\hfill\sq}
\newcommand{\bp}{\begin{proof}}
\newcommand{\ep}{\end{proof}\par\vspace{10pt}\noindent}
\begin{document}

\title{On ``full'' twisted Poincar\'e symmetry and
QFT on Moyal-Weyl spaces}

\author{   Gaetano Fiore${}^{1,2,3}$, 
Julius Wess${}^{3,4,5}$  \\\\
        %  \and
$^{1}$ Dip. di Matematica e Applicazioni, Universit\`a ``Federico II''\\
   V. Claudio 21, 80125 Napoli, Italy\\         %\and
$^{2}$         I.N.F.N., Sez. di Napoli,
        Complesso MSA, V. Cintia, 80126 Napoli, Italy\\
        \and
 ${}^{3}$ Arnold Sommerfeld Center for Theoretical Physics\\
Universit\"at M\"unchen,
Theresienstr.\ 37, 80333 M\"unchen, Germany\\
  %   \and
${}^{4}$ Max-Planck-Institut f\"ur Physik\\
        F\"ohringer Ring 6, 80805 M\"unchen, Germany\\
    % \and
${}^{5}$ Universit\"at Hamburg, II Institut f\"ur Theoretische Physik\\
and DESY, Luruper Chaussee 149, 22761 Hamburg, Germany
}
\date{}

\maketitle \abstract{We explore some general
consequences of a proper, {\it full}
enforcement   of  the ``twisted Poincar\'e'' covariance
%(i.e. covariance under a deformed  Poincar\'e Hopf algebra obtained by twisting)
of Chaichian {\it et al} \cite{ChaKulNisTur04}, Wess \cite{Wes04}, 
Koch {\it et al} \cite{KocTso04}, Oeckl \cite{Oec00} upon
many-particle quantum mechanics and field quantization on a Moyal-Weyl
noncommutative space(time). This entails the associated  braided
tensor product  with an involutive braiding (or $\star$-tensor
product in the parlance of Aschieri {\it et al}
\cite{AscBloDimMeySchWes05,AscDimMeyWes06}) prescription
for any coordinates pair of $x,y$ generating two different copies of the 
space(time);
the associated nontrivial commutation relations between them imply that $x-y$
is central and its Poincar\'e transformation properties remain undeformed.
As a consequence, in QFT (even with space-time noncommutativity) one can reproduce notions (like
space-like separation,  time- and normal-ordering, Wightman or Green's
functions, etc), impose constraints (Wightman
axioms), and construct free or interacting theories which essentially coincide
with  the undeformed ones, since the only observable quantities involve
coordinate differences.
In other words, one may thus well realize QM and QFT's where the effect of
space(time) noncommutativity amounts to a practically unobservable common
noncommutative translation of all reference frames.

}

\vskip1cm
\noindent
- Preprint 06-54 Dip. Matematica e Applicazioni, Universit\`a di Napoli;\\
\noindent
- DSF/2-2007

\newpage

\section{Introduction: Moyal-Weyl spaces, twisted Poincar\'e ``group'' and QFT}

In the last decade a broad attention has been devoted to the construction
of Quantum Field Theories (QFT) on  the perhaps simplest examples
of noncommutative spaces, the socalled Moyal-Weyl spaces. These are
characterized by  coordinates $\hat x^\mu$ fulfilling
the commutation relations
\be
[\hat x^\mu,\hat x^\nu]=i\theta^{\mu\nu}\ ,\label{cr}
\ee
where $\theta^{\mu\nu}$ is a constant real antisymmetric
matrix. The $\theta^{\mu\nu}=0$ limit is the undeformed algebra $\A$
generated by commuting coordinates $x^{\mu}$.
For the sake of definiteness we shall suppose
(with the exception of Section \ref{QM})  $\mu=0,1,2,3$  and endow
the space with the ordinary Minkowski metric $\eta_{\mu\nu}$, to
obtain a deformation of the $3+1$-dimensional Minkowski spacetime. As
$\theta^{\mu\nu}$ is not an isotropic tensor, relations (\ref{cr}) are  not
covariant (i.e. not form-invariant) under Lorentz transformations of the
reference frame (although they are  invariant under translations).

The unital  algebra
$\hA$ generated by these $\hat x^{\mu}$ is isomorphic
to the one $\A_{\theta}$ which is obtained
by endowing the vector space underlying $\A$ (extended over the formal power series in $\theta^{\mu\nu}$)
with a deformed product, the $\star$-product, which can be
formally defined by
\be
a\star b:=(\bF^{(1)}\trc a)  (\bF^{(2)}\trc b).            \label{starprod}
\ee
For typographical convenience
we have denoted by $\bF\equiv\F^{-1}$ the inverse of the socalled twist $\F$.
It (and therefore also the associated isomorphism $\phi:\hA\to\A_{\theta}$)  is
not uniquely determined, but what follows does not depend on the specific
choice of $\bF$. The simplest is
\be
\bF\equiv \bF^{(1)}\ot\bF^{(2)}:=
\mbox{exp}\left(-\frac i2\theta^{\mu\nu}P_{\mu}\ot P_{\nu}\right).
                                                   \label{twist}
\ee
$P_{\mu}$ denote the generators of translations, and $\trc$ in general
denotes the action of the Universal Enveloping algebra (UEA) $U{\cal P}$ of
the Poincar\'e Lie algebra ${\cal P}$ (on $\A$ this amounts to the action of
the corresponding algebra of differential operators, e.g.   $P_{\mu}$ can be
identified with $i\partial_{\mu}:= i\partial/\partial x^{\mu}$).
 In the second expression and in (\ref{starprod}) we have used a Sweedler notation
with suppressed summation index: $\bF^{(1)}\!\ot\!\bF^{(2)}$ stands in fact for a
(infinite) sum $\sum_I\bF_I^{(1)}\!\ot\!\bF_I^{(2)}$.
Relation (\ref{starprod}) with the specific choice (\ref{twist}) of the
twist gives  in particular
$$
\hat x^\mu \hat x^\nu \stackrel{\phi}{\longrightarrow} x^\mu\!\star\!
x^\nu=x^\mu x^\nu+ i\theta^{\mu\nu}/2.
$$
As a result,
$x^\mu\!\star\! x^\nu-x^\nu\!\star\! x^\mu=i\theta^{\mu\nu}$, i.e. again
(\ref{cr}), as claimed.
One advantage of working with  $\A_{\theta}$ instead od $\hA$
is that integration
over the original commutative space can be used also on the noncommutative
one without loosing its properties (in particular Stoke's theorem). In addition,
\be
\int d^4x \, a\star b=\int d^4x \, ab
\ee
for any regular $a,b$ functions in the vector space underlying $\A$ vanishing
sufficiently fast at infinity. 
The definition (\ref{starprod}-\ref{twist}) involves a power series in 
$\theta^{\mu\nu}$ and for the moment should be regarded as formal: it can be applied to a much larger domain if $\bF$ is rather
realized as the integral operator, as we shall explain in (\ref{IntForm}).

Different (obviously not Lorentz-covariant) approaches  to quantization of
field theory on Moyal-Weyl spaces  have been proposed
(see \cite{DopFreRob95,Fil96}, \cite{DouNek01,
Sza03} and references therein).
New complications appear, like non-unitarity \cite{GomMeh00},  violation of
causality \cite{SeiSusTou00,BozFisGroPitPutSchWul03},
UV-IR mixing of divergences \cite{MinRaaSei00}
and subsequent non-renormalizability, alleged change of statistics, etc.  Some
of these problems, like non-unitarity  \cite{BahDopFrePia02}, or the very
occurrence  of divergences \cite{BahDopFrePia03-05},  may be
due simply to  naive (and unjustified) applications of commutative QFT rules
(path-integral methods, Feynman diagrams, etc) and could disappear adopting a
sounder   field-operator approach.

In Ref.  \cite{ChaKulNisTur04,Wes04,KocTso04} it has been recognized that
the commutation relations of $\hA\sim \A_{\theta}$ are in fact covariant
under a deformed version of the Poincar\'e group, namely the
triangular noncocommutative Hopf $*$-algebra $H$ obtained from
 $U{\cal P}$ by ``twisting''
\cite{Dri83} with $\F$ (this result had been in fact anticipated in terms of corepresentations of the dual Hopf algebra in section 4.4.1 of \cite{Oec00}. 
For a general introduction to the twist see e.g. \cite{ChaPre94}).
This means that (up to possible  isomorphisms) the algebra structure and the
counit $\varepsilon$ of $U{\cal P},H$ (extended over the formal power series 
in $\theta^{\mu\nu}$)
are the same, but the coproduct is changed through the similarity
transformation \be
\Delta(g)\equiv g_{(1)}\ot g_{(2)}\quad\longrightarrow\quad
\hat\Delta(g)=\F\Delta(g)\F^{-1}\equiv g_{(\hat 1)}\ot g_{(\hat 2)},
\qquad\qquad g\!\in\! H\!=\! U{\cal P}                        \label{coproductn}
\ee
(at the rhs's we have again used Sweedler notation with
suppressed summation indices), and the antipode $S$ accordingly. 
A straightforward computation  gives
$$
\hat\Delta (P_\mu)=P_\mu\!\ot\!\1\!+\!\1\!\ot\! P_\mu=
\Delta(P_\mu),\qquad\quad\hat\Delta (M_\omega)=M_\omega\!\ot\!\1\!+\!\1\!\ot\! M_\omega+
P[\omega,\theta]\!\ot\! P\neq\Delta (M_\omega),
$$
where we have set $M_\omega\!:=\!\omega^{\mu\nu}M_{\mu\nu}$ and used
a row-by-column matrix product on the right.
The left identity shows that the Hopf $P$-subalgebra remains undeformed and equivalent to the
abelian translation group  ${\cal T}\sim\b{R}^4$.
Denoting by $\trc,\tr$ the (say, left) actions of 
$U{\cal P},H$, they coincide on first degree polynomials in 
$x^\nu,\hat x^\nu$,
\be
P_{\mu}\trc x^{\rho}=i\delta^{\rho}_{\mu}=P_{\mu}\tr \hat x^{\rho},
\qquad\quad M_\omega\trc x^{\rho}=2i(x\omega)^\rho,
\qquad\quad M_\omega\tr \hat x^{\rho}=2i(\hat x\omega)^\rho,
   \label{Px}
\ee
and more generally on irreps (irreducible representations); 
as noted in \cite{ChaKulNisTur04}, 
this yields the same
classification of elementary particles as unitary irreps of ${\cal P}$. But
$\trc,\tr$ differ on products of coordinates, and more generally on tensor products of representations, as $\trc$ is extended by the rule
$g\trc\! (ab)\! =\!\big( g_{(1)}\!\trc a\big) \!\big( g_{(2)}\!\trc
b\big)$ involving $\Delta(g)$ (the rule reduces
to the usual Leibniz rule for $g=P_\mu,M_{\mu\nu}$), 
whereas $\tr$ is extended on products of elements in both
$\hA,\A_{\theta}$ by the rule
\be
g\tr (\hat a\hat b) =\big( g_{(\hat 1)}\tr \hat a\big)
\big( g_{(\hat 2)}\tr \hat b\big)\qquad\Leftrightarrow\qquad
g\tr (a\!\star\! b) =\big( g_{(\hat 1)}\tr
a\big)\star \big( g_{(\hat 2)}\tr b\big),
\label{Leibnizn}
\ee
which respects the commutation relations (\ref{cr}), making $\hA,\A_{\theta}$
isomorphic $H$-module algebras; this deforms in particular the Leibniz rule 
of $M_{\mu\nu}$ (but not of $P_\mu$). 

How to implement this twisted Poincar\'e covariance in QFT is subject
of debate and different proposals
\cite{ChaNisTur06,ChaPreTur05,Tur06,BalManPinVai05,
BalPinQur05,BalGovManPinQurVai06,BuKimLeeVacYee06,Zah06,LizVaiVit06,Abe06},
 two main issues being whether
one should: {\it a)} take the $\star$-product of fields at different spacetime points;
{\it b)}  deform the canonical commutation relations (CCR)
of creation and annihilation operators $a,a^\dagger$ for free fields.

The aim of this work is to point out that a proper %, ``full''
enforcement of twisted Poincar\'e covariance answers affirmatively to {\it a)}
and brings a radical simplification to the framework, in that all
coordinate differences become $\star$-central, i.e. central w.r.t. the
$\star$-product (section \ref{manyx}). We first explore (section \ref{QM})
some consequences of the latter fact in $n$-particle Quantum Mechanics
(QM): we find that twisted Galilei covariance is compatible with
Bose or Fermi statistics and that the dynamics of an isolated system
of $n$-particles is the same as its counterpart on commutative space. As
for QFT, which we treat in field-operator approach, we sketch the general
consequences of (slightly adapted) Wightman axioms in section \ref{QFT}, show in section \ref{Free} that the latter can be satisfied by free (for simplicity scalar) fields if we also suitably deform the CCR of the $a,a^\dagger$'s
so that the $\star$-commutator of the
fields is equal to the undeformed counterpart,
show in section \ref{Interact} that then the time-ordered 
perturbative computation of Green functions of a scalar
$\varphi^{\star n}$ interacting theory gives the same results as the
undeformed theory.
%The latter is a conclusion similar to the one of \cite{BalPinQur05} about
%the $S$-matrix remaining undeformed, but is in fact much stronger.
In other words, we end up in this way with twisted Poincar\'e
covariant QFT's which are  physically equivalent
to their counterparts on commutative Minkowski space, with the obvious
consequence that the above-mentioned complications will disappear.
In Section \ref{outlook} we draw some conclusions and briefly comment
on the alternatives implying violation of the cluster property
by the Wightman functions.

\section{The action of the twisted Poincar\'e ``group'' on several
spacetime variables}
\label{manyx}

Dealing with $n$-point (Green's, or Wightman's, etc) functions in QFT
requires $n$ sets of noncommutative Minkowski spacetime coordinates $\hat x^{\mu}_i$,
$i=1,...,n$,  of type (\ref{cr}).
Similarly, dealing with $n$-particle QM
requires $n$ sets of noncommutative Euclidean space coordinates $\hat
x^{\mu}_i$, (one for each particle)  of type (\ref{cr}).

Our starting, basic observation
%in the present paper
is that to consistently adopt the
viewpoint of twisted Poincar\'e covariance {\it one should require that also the
larger algebra $\hA^n$ generated by them is a $H$-module algebra},
meaning in particular that within the latter (\ref{Leibnizn}) still holds.
This is also the philosophy adopted in Ref. \cite{AscDimMeyWes06}.
To this end one {\it cannot} adopt as
$\hA^n$  the tensor product algebra of $n$ copies of $\hA$, or equivalently
assume trivial commutation relations
$$
[\hat x_i^{\mu},\hat x_j^{\nu}]=0 \qquad\qquad i\neq j,
$$
as done e.g. in
\cite{Fra05,BalGovManPinQurVai06}, because the latter are
incompatible with (\ref{Leibnizn}) by the non-cocommutativity of $\hat\Delta$
(this can be checked e.g. by letting the Lorentz generators $M^{\rho\sigma}$
act on both sides).
In fact it is a basic property of quasitriangular Hopf algebra theory
(see e.g. %\cite{Abe?,Dri86,ChaPre,Maj95},
\cite{Maj95}) that one has to adopt as
$\hA^n$ rather the deformation of  the tensor product algebra,
usually called {\it braided tensor product algebra},
dictated by the  quasitriangular structure  $\R$ of $H$.
Given two left $H$-module algebras $\hat M,\hat M'$ the  braided
tensor product algebra $\hat M\uot \hat M'$  is still $\hat M\ot \hat M'$
as a vector space, but is characterized by the product
\be
(\hat m\uot\hat  m')\,\cdot\,(\hat n\uot\hat  n')=\hat m(\R^{(2)}\tr  \hat n
)\uot (\R^{(1)}\tr\hat  m') \hat n', \label{primo}
\ee
where we have again used a Sweedler notation with suppressed
summation index: $\R\equiv \R^{(1)}\ot \R^{(2)}$ stands in fact for a
(infinite) sum $\sum_I\R_I^{(1)}\!\ot\!\R_I^{(2)}$.
In the present
case $\R=\F_{21}\F^{-1}=(\F)^{-2}$ is even triangular, i.e.
$\R\R_{21}=\1\ot\1$, implying that
these rules are symmetric w.r.t. to the exchange of $\hat M,\hat M'$,
or equivalently the braiding coincides with the ordinary flip up to
a similarity transformation.
If $\hat M,\hat M'$ are $H$-module algebras,
deformations of two $U{\cal P}$-module algebras $M,M'$, so that
the isomorphisms $\hat
M\!\sim\! M_{\theta}$,  $\hat
M'\!\sim\! M'_{\theta}$ hold, the braided tensor product
(\ref{primo}) is isomorphic to the $\star$-tensor product $\ots$ of
\cite{AscDimMeyWes06}, which is defined  by setting
for any $m\in M_{\theta}$, $m'\in M'_{\theta}$
\be
 m\ots  m'=(\bF^{(1)}\trc m) \ot (\bF^{(2)}\trc m').    \label{startensprod}
\ee
That this is the `right' deformation of the tensor product follows also from
the observation that {\it this is nothing but the extension of the
$\star$-product law (\ref{starprod}) to the whole tensor product algebra $M\ot
M'$}, in the sense \be
m\ots  m'=(m\ot \1)\star (\1\ot m').      \label{startensprod'}
\ee
If $\hat M,\hat M'$ are unital (\ref{primo}) reduces to the ordinary tensor
algebra rule if either $\hat m'\!=\!\1$ or $\hat n\!=\!\1$, as
$\varepsilon( \R^{(1)})\R^{(2)}\!=\!\varepsilon( \R^{(2)})\R^{(1)}\!=\!\1$.
As for ordinary tensor product algebras, because of the trivial algebra  isomorphisms
$\1\uot \hat M'\sim \hat M'$, $\hat M\uot\1\sim \hat M$, one can
simplify the notation by dropping the
units, i.e. denote $\hat m\uot \1$ and $\1\uot \hat m'$ resp. by $\hat m,\hat m'$,
whereby the only novelty of (\ref{primo}) remains concentrated in the
nontrivial ``cross'' commutation relation
$$
\hat m'\hat n= (\R^{(2)}\tr  \hat n )\,(\R^{(1)}\tr\hat  m').
%\label{Rcomrel}
$$
Similarly, we can simplify the notation denoting the sides of
(\ref{startensprod'}) as $m\star  m'$ and replacing the previous relation
by $m'\star n=(\R^{(2)}\tr  n )\star(\R^{(1)}\tr m')$.

\medskip
Choosing as $\hat M,\hat M'$ two copies of the  $*$-algebra of functions
$\hA$  on the Moyal-Weyl noncommutative space,  calling $\hat x,\hat y$
the respective sets of coordinates, and noting that the action
of the translation generators on the coordinates is given by
$$
P_{\mu}\tr \hat x^{\nu}=P_{\mu}\tr \hat y^{\nu}=i\delta^{\nu}_{\mu}
%   \label{Px}
$$
we find
$$
\hat x^{\mu}\hat y^{\nu}= (\R^{(2)}\tr  \hat y^{\nu} )\,(\R^{(1)}\tr
\hat x^{\mu})
=\hat y^{\nu}\hat x^{\mu}+i\theta^{\mu\nu}.
$$
These are also automatically compatible with the $*$-structure
(a straightforward check, beside a consequence of $\R^{*\ot
*}=\R_{21}=\R^{-1}$), and with setting $\hat x=\hat y$.
More generally, applying the above rule iteratively, the braided tensor
product of $n$ copies of $\hA$ and the
$\star$-tensor product of $n$ copies of $\A_\theta$  will be isomorphic
$H$-module $*$-algebras $\hA^n,\A^n_\theta$ respectively generated by real
variables $\hat x^{\mu}_i$ and $x^{\mu}_i$, $i=1,2,...,n$, fulfilling the
commutation relations \be
[\hat x^{\mu}_i,\hat x^{\nu}_j]=i\theta^{\mu\nu}
\quad\qquad\Leftrightarrow\quad\qquad [ x^{\mu}_i\stackrel{\star},
x^{\nu}_j]=i\theta^{\mu\nu}.     \label{summary}
\ee
This formula summarizes all the commutation relations defining
$\hA^n\!\sim\!\A^n_\theta$:
for $i\!=\!j$ these are the defining commutation relations of the $i$-th copy, for
$i\!\neq\! j$ these are consequences of the braided tensor (or $\star$-tensor)
product between the  $i$-th and the $j$=th copy. 
Summing up,
the algebra $\A^n_\theta$ is
obtained by endowing the vector space underlying the $n$-fold tensor product
$\A^n$ of $\A$
with a new product, the $\star$-product, related to the product in $\A^n$ by
formula (\ref{starprod}) for any $a,b\in\A^n$.
This encodes both the
usual $\star$-product within each copy of $\A$, and
the $\star-$tensor product of
\cite{AscBloDimMeySchWes05,AscDimMeyWes06}. More explicitly, on 
analytic functions $a,b$ (\ref{starprod}) reads
\be
a(x_i)\star b(x_j)= \exp\left(\frac i2\partial_{x_i}\theta\partial_{x_j}\right)
a(x_i) b(x_j),                               \label{explstarprod}
\ee
and must be followed by the indentification $x_i\!=\!x_j$ 
{\it after} the action of the bi-pseudodiffer\-ential operator
$\exp[\frac i2\partial_{x_i}\theta\partial_{x_j}]$ if $i\!=\!j$.

Strictly speaking, the definitions (\ref{starprod}-\ref{twist}) 
or (\ref{explstarprod}) make sense if 
we choose $a,b$ in a suitable subspace  $\A'\!\subset \!\A$
ensuring that the involved power series in $\theta^{\mu\nu}$ is termwise well-defined and converges. One such subspace can be looked for within the space of (analytic) functions 
that are the Fourier transforms $\hat g$ of functions $g$ with compact support. The determination of the largest possible $\A'$
is a delicate issue, about which little is known (see \cite{EstGraVar89}
and references therein).
Anyway for field-theoretic purposes it would not be enough to work
with $\A'$, and it is much better to {\it define} the 
 $\star$-product as the integral with a non-local kernel 
\be
a(x_i)\star b(x_j)= \int\! d^4h\!\int\!  d^4k\:\,
e^{i\left(h\cdot x_i+k\cdot x_j-\frac{h\theta k}2\right)}\check a(h) \check b(k),
                                                       \label{IntForm}
\ee
where  $\check{}$ denotes the antiFourier
transform. This is well-defined
if $a,b\in L^1(\b{R}^4)\cap \widehat{L^1(\b{R}^4)}$, can be defined even
if $a,b$ are distributions, and is designed
so as to have the series (\ref{starprod}) as a formal power expansion; 
see \cite{EstGraVar89} for the conditions under which the latter is in fact an asymptotic expansion.  More generally, one should adopt 
as proper definition of the action
of $\F,\bF$ and of derived operators like $(\Delta\!\ot\!\id) \F$
the corresponding non-local integral operators. They also fulfill
the cocycle condition
$\F_{12}(\Delta\!\ot\!\id) \F=\F_{23}(\id\!\ot\!\Delta) \F$, 
ensuring the associativity of the $\star$-product.

\bigskip
\noindent
We now define an alternative set of real generators of $\A^n_\theta$
(or, correspondingly, of $\hA^n$):
\be
\xi^{\mu}_i:=  x^{\mu}_{i\!+\!1}\!-\! x^{\mu}_i,\quad         i=1,...,n\!-\!1,
\qquad\qquad\qquad
 X^{\mu}:=\sum\limits_{i=1}^na_i x^{\mu}_i
\ee
where $a_i$ are real numbers  such that $\sum_ia_i=1$
(in particular one could choose $ X^{\mu}= x^{\mu}_j$, for some
special $j$). It is immediate to verify that:

\begin{enumerate}

 \item  All $\xi^{\mu}_i$ are invariant under translations,
(whereas $ X^{\mu}$ is not):
\be
P_{\mu}\tr  \xi^{\nu}_i=0, \qquad\qquad
P_{\mu}\tr   X^{\nu}=i\delta^{\nu}_{\mu}.               \label{trivialPz}
\ee

\item
$ X^{\mu}$  generate a copy $\A_{\theta,X}$ of Moyal-Weyl
noncommutative space,  whereas the $\star$-product with $\xi^{\mu}_i$
(or any function thereof) reduces to the ordinary product
\be
\xi^{\mu}_i\star b=\xi^{\mu}_i b=b\star \xi^{\mu}_i\qquad\qquad\qquad
b\!\in\!\A_{\theta}^n,                          \label{startrivial}
\ee
implying that the  $\xi^{\mu}_i$ are $\star$-central in $\A^n_\theta$
(i.e. $\star$-commute with everything),
\be
[\xi^{\nu}_i\stackrel{\star},\A^n_\theta]=0.
\ee
Thus the  central $*$-subalgebra $\A_{\theta,\xi}^{n\!-\!1}$  generated by the
$\xi^{\mu}_i$ reduces to the ordinary tensor product algebra of $n\!-\!1$
copies of the undeformed $\A$ [because of the trivial action
(\ref{trivialPz}) of the $P_\mu$ contained in the twist
$\F=\mbox{exp}\left(\frac i2\theta^{\mu\nu}P_{\mu}\ot P_{\nu}\right)$
and in $\R=\F^{-2}$ on the tensor factors], whereas $\A^n_\theta$ reduces to
the tensor product  algebra
$\A^n_\theta=\A_{\theta,\xi}^{n\!-\!1}\ot\A_{\theta,X}$. Moreover, the
$\xi^{\mu}_i$ have the same spectral decomposition on the whole $\b{R}$ as
classical variables $\xi^{\mu}$; in particular, 0 is in their spectrum.

\item  $\A_{\theta,\xi}^{n\!-\!1},\A_{\theta,X}$ are actually $H$-module
subalgebras, and
\be
g\tr ( a\!\star\! b) \!=\!\left( g_{(1)}\trc a\right)\!\star\!
\left( g_{(2)}\tr b\right) ,    \qquad\qquad    a\!\in\!
\A_{\theta,\xi}^{n\!-\!1},\quad b\!\in\!\A_{\theta}^n,\quad
g\in H,                    \label{UndefLeibniz}
\ee
implying in particular $g\tr a\!=\!g\trc a$, i.e. {\it on
$\A_{\theta,\xi}^{n\!-\!1}$ the $H$-action is undeformed}. In fact the Leibniz
rule reduces to the undeformed one whenever a twist leg acts on $a$, again
because of the trivial action  (\ref{trivialPz})$_1$ of the $P_\mu$'s contained
in $\bF$. The previous relation holds also without the two $\star$-products,
by (\ref{startrivial}).

\end{enumerate}
Summing up, any coordinate difference like $\xi^{\mu}_i$ can be
treated as a classical, commutative variable. Any $ x^{\mu}_i$ is
a combination of $n\!-\!1$ $\star$-commutative variables $\xi^{\mu}_i$ and 1
$\star$-noncommutative one $X^{\mu}$; or equivalently can be obtained from
the zero 4-vector and $n\!-\!1$ $\star$-commutative 4-vectors   by
the global ``noncommutative translation''  $X$, e.g. if $X:= x_1$
then
$$
 x_i=\sum\limits_{j=1}^{i-1}\xi_j +  X.
$$
Of course, all the previous statements [with the exception of
(\ref{startrivial})] can be formulated
in the isomorphic setting removing all $\star$'s, putting a
$\hat{}$ over any coordinate and replacing
$\A_\theta,\trc_\star,\A^n_\theta,\A_{\theta,\xi}^{n\!-\!1},\A_{\theta,X}$
with the isomorphic
objects $\hA,\tr,\hA^n,\hA_{\xi}^{n\!-\!1},\hA_X$
The result for $\hat X$ is like the ``quantum shift operator''  of
\cite{ChaNisTur06}.

\medskip
{\bf Remark 1.}
One immediate consequence is that on any irreducible representation
$\star$-multiplication by a
spacetime coordinate difference $x\!-\!y$ equals multiplication by $x\!-\!y$,
which is
either a space-like, or a null, or a time-like $4$-vector, in the usual sense.

\medskip
{\bf Remark 2.} Relation (\ref{UndefLeibniz}) holds also for an
infinitesimal  general coordinate transformation, i.e. if $g$ is an element
of the (deformed) U.E.A. $U\Xi_{\star}$
\cite{AscBloDimMeySchWes05,AscDimMeyWes06} of the Lie algebra
of general vector fields on the Moyal-Weyl NC space.

\medskip
We recall that the differential calculus over $\b{R}^n$ remains
unchanged under deformation of this space into a Moyal Weyl NC space.
This is true also if we consider the differential calculus
on the larger algebra  $\A^n_\theta$ (or the isomorphic $\hA^n$),
and follows again from (\ref{starprod}), (\ref{twist}) and the fact that
$P_\mu$ have trivial action on the derivatives.
Explicitly,
\be
%\frac{\partial}{\partial \hat x^{\mu}_i}\hat
%x^{\nu}_j=\delta^\nu_\mu\delta^i_j+ \hat x^{\nu}_j\frac{\partial}{\partial
%\hat x^{\mu}_i} \qquad\Leftrightarrow \qquad
\partial_{
x^{\mu}_i}\star x^{\nu}_j=\delta^\nu_\mu\delta^i_j+
x^{\nu}_j\star \partial_{x^{\mu}_i}
 \qquad\qquad\qquad
\left[\partial_{x^{\mu}_i}\stackrel{\star},\partial_{x^{\nu}_j}\right]=0
\label{derivatives}
\ee
with self-explaining notation. Since the presence of
the $\star$ product has no effect on the action of the derivatives on
$\A^n_\theta$, in the sequel we shall drop it.

\medskip
Given two sets $x,y$ of coordinates,
integrating over some $x^\mu$ both sides of the identity
$$
g(y) \star f(x)=(\R^{(2)}\tr  f(x) )\star(\R^{(1)}\tr g(y) )
=\mbox{exp}\left(-i\theta^{\mu\nu}\partial_{x^\mu}\partial_{y^\nu}\right)
f(x) \star g(y)
$$
we see that any integration $\int\! dx^\mu$ commutes with $g(y)\star$
if $f$ rapidly decreases at infinity;  in fact, if we define the 
$\star$-product by the integral (\ref{IntForm}) we realize that 
\be
\int\! dx^\mu\, g(y) \star f(x)= g(y) \star \int\! dx^\mu \,f(x)
                                                 \label{starIntCR}
\ee
is true also for $f,g\in L^1(\b{R}^4)\cap \widehat{L^1(\b{R}^4)}$ or even some
 distributions, as on commutative space [of course, since the 
rhs(\ref{starIntCR}) is independent of $x^\mu$,  terms
with $\theta^{\mu\nu}\partial_{y^\nu}$ will be ineffective
and disappear, as if it were $\theta^{\mu\nu}\!=\!0$
for all $\nu$]. 
Therefore, for our purposes we can consider integration
over any set of coordinates as an operation commuting with $\star$-products.

\section{General consequences for many-particle QM}
\label{QM}

In configuration space the Hamiltonian of an isolated system of $n$ non-relativistic (for simplicity spinless) particles 
\be
\Ha=\Ha_0+\sum\limits_{i<j} V_{ij}(|{\bf x}_i-{\bf x}_j|)
\qquad \qquad\qquad
\Ha_0:=-\sum\limits_{i=1}^{n}\frac{\hbar^2}{2m_i}\nabla^2_{{\bf x}_i}
\ee
involves only derivatives and relative coordinates $\xi$. Denoting as ${\bf X}$
the coordinates of the center of mass, as $M$ the total mass of the system, the
kinetic part $\Ha_0$ can be written as the sum of $-\hbar^2\nabla^2_{\bf X}/2M$
and a second order differential operator in the $\xi$-derivatives only.
As a consequence, the dynamics of the center of mass is free. This means that an orthogonal basis of eigenfunctions of
$\Ha$ is   $\{\exp(i{\bf k}\!\cdot \!{\bf X})\psi_j(\xi)\}$, where $\psi_j$ are
eigenfunctions of the rest Hamiltonian
$\Ha_\xi:=\Ha\!+\!\hbar^2\nabla^2_{\bf X}/2M$, depending on the $\xi$ and
$\xi$-derivatives only.

Going to the noncommutative Euclidean space (the time remaining
commutative) brings no change: the deformed
Hamiltonian $\Ha_\star\equiv\Ha\star$ can be still split into a free part
$-\hbar^2\nabla_{\bf X}^2/2M\star$  for the center-of-mass degrees of freedom
and an interacting
part $\Ha_{\xi}\star$ depending only on the relative coordinates, and both parts
act on the vector space underlying both $\A^n_\theta$ and $\A^n$ (and therefore
also on the subspace consisting of square-integrable
wave-functions) exactly as their undeformed counterparts, implying that
$\{\exp(i{\bf k}\!\cdot\! {\bf X})\psi_J(\xi)\}$ is also an orthonormal basis of
eigenfunctions of $\Ha_\star$ with the same eigenvalues. As a result, the
deformed dynamics coincides with the undeformed one.

\medskip
Assume now that the  particles are identical. If the space
is commutative, a  wave-function
$\Psi({\bf x}_1,...,{\bf x}_n)$ completely (anti)symmetric under
particles' permutations can be decomposed as
$\Psi=\sum_{IJ}\Psi_{IJ}\phi_I\chi_J$ in any tensor product basis
$\{\phi_I({\bf X})\chi_J(\xi)\}$, where $\chi_J(\xi)$ are completely (anti)symmetric
[$\phi_I({\bf X})$ are automatically completely symmetric]. 
These symmetries are preserved by the dynamical evolution, since this is ruled
by the completely symmetric evolution operator 
$U(t\!-\!t_0)=\exp[-\frac i{\hbar} \Ha' (t\!-\!t_0)]$, where
$\Ha'=\Ha+\sum_i V_e({\bf x}_i)$ is the total Hamiltonian with
$V_{ij}\equiv V$  and $V_e$ the external potential (if the system 
is not isolated). 
For the same reason this is true both in the Schr\"odinger and in the 
Heisenberg picture, which are related by the unitary transformation
$U(t\!-\!t_0)$, and also in the interaction picture, which is related to the
Schr\"odinger by the completely symmetric evolution operator 
$U_0(t\!-\!t_0)=\exp[-\frac i{\hbar} \Ha_0 (t\!-\!t_0)]$.
All  the corresponding deformed statements remain true, as
$\Ha'_\star\equiv\Ha' \star$ and 
$\Ha_{0\star}\equiv\Ha_0 \star$ are also completely symmetric.

The action of the Galilei 
Lie algebra ${\cal G}$\footnote{We recall that ${\cal G}$ is generated by
$H_0$ (kinetic term in the Hamiltonian: generates time translations of
a free system),
$m$ (mass: is central), 
$P^a$ (momentum components: generate space translations), 
$L^a$ (angular momentum components: generate rotations), 
$K^a$ (generate boosts), 
with $a=1,2,3$, 
where the only nontrivial commutation relations are
\be
\ba{lll}
[K^a,P^b]=im \hbar\delta^{ab}, \qquad &[K^a,H_0]=i\hbar P^a, & \\[8pt]
[L^a,L^b ]=i\epsilon^{abc}\hbar L^c,\qquad &[L^a,P^b]=i\epsilon^{abc}\hbar P^c,
\qquad &[L^a,K^b]=i\epsilon^{abc}\hbar K^c.
\ea
\ee
The generators are realized as the
differential operators
$H_0=-\hbar \nabla^2/2m$, $P^a=-i\hbar \partial^a$,  
$L^a=-i\hbar \epsilon^{abc}x^b\partial^c$, $K^a=mx^a+i\hbar t\partial^a$
in the configuration space of each single particle. Hence the observable
$K^a+tP^a$ gives the mass times the space coordinate $x^a$ of the particle.
The coproducts are defined by the fact that these
generators are primitive. The coproducts of 
$m,H_0, P^a,L^a$ respectively give 
the addition laws for the total mass, the total kinetic energy, 
the total momentum and the total angular momentum of the system, whereas the coproduct of  $K^a+tP^a$
gives the total mass times the space coordinate $X^a$ of the center of mass
of the system.}, and therefore also
of its universal enveloping algebra $U{\cal G}$,  maps $\A_X\to\A_X$,
$\A_{\xi}^{n\!-\!1}\to\A_{\xi}^{n\!-\!1}$ preserving these
complete (anti)symmetries, hence amounts to a change of the coefficients
$\Psi_{IJ}$.
Interpreting $\Psi,\phi_I({\bf X}),\chi_J(\xi)$ as elements
respectively of $\A^n_\theta,\A_{\theta,X},A_{\theta,\xi}^{n\!-\!1}$,
the same will be true of the action of the twisted
Galilei U.E.A. $H$, as the latter maps
$\A_{\theta,X}\to\A_{\theta,X}$,
$A_{\theta,\xi}^{n\!-\!1}\to A_{\theta,\xi}^{n\!-\!1}$, by
(\ref{UndefLeibniz}).  Therefore  there is no incompatibility between the
standard complete (anti)symmetry conditions on a wave-function
$\Psi({\bf x}_1,...,{\bf x}_n)$  and the action of $H$.  
Consequently, the standard Bose,
Fermi (and similarly anyon, in 2 space dimensions) statistics are compatible
with twisted Galilei symmetry (in first quantization). This agrees with the
general (and physically reassuring!) results of Ref. \cite{FioSch96}, where it
was shown (by a unitary equivalence in a $n$-particle, abstract Hilbert space
formalism) that covariance under a noncocommutative Hopf algebra obtained by
twisting from a cocommutative one is compatible with usual statistics.

\section{General consequences for QFT}
\label{QFT}

In field-operator approach quantization of fields on Minkowski space
obeys a set of general conditions, the Wightman axioms \cite{StrWig63}, which
(as done e.g. in Ref. \cite{Stro04}) can be
divided into a subset (in the sequel labelled by {\bf QM})  encoding
the quantum mechanical interpretation of the theory, its symmetry under
space-time translations and stability, and a
subset (in the sequel labelled by {\bf R}) encoding the relativistic
properties (full Lorentz-covariance and locality).
We now try to translate this into a field quantization
procedure on a Moyal-Weyl noncommutative space keeping the QM
conditions, ``fully'' twisting Poincar\'e-covariance and being ready to
weaken locality if necessary.

%only changes that products should be $\star$-products and that
%Poincar\'e covariance should be twisted.

%%%%%%%%%

\medskip
\noindent
\vspace{1mm} {\bf QM1.} ({\bf Hilbert space structure})
The states are described by vectors of a (separable) Hilbert space $\H$.

\medskip
\vspace{1mm}
\noindent
{\bf QM2.} ({\bf Energy-momentum spectral condition}) The group of
space-time translations ${\cal T}\sim\b{R}^4$ is a symmetry of the theory and
is represented  on $\H$ by strongly continuous unitary operators
$U(a), \, a \in \b{R}^4$: the fields transform according to (\ref{transf})
with unit $A,U(A), \Lambda(A)$.
The spectrum of the generators $P_\mu$ is contained in the closed
{\em forward cone} $\overline{V}_+ = \{p_\mu: p^2 \geq 0, \,p_0 \geq 0
\}$. There is a {\em vacuum state} $\Psi_0$, with the property of
being the unique %(twisted)
Poincar\'e invariant state
({\em uniqueness of the vacuum}).

\vspace{1mm}
\noindent
{\bf QM3.} ({\bf Field operators}) The theory is formulated in terms of
fields (in the Heisenberg representation)
$\varphi^\alpha(x)$, $ \alpha\!= \!1, ...N$, that are operator
(on $\H$) valued tempered distributions on Minkowski space, with $\Psi_0$ a
{\em cyclic} vector for the fields, i.e. by applying polynomials of the
(smeared) fields to $\Psi_0$ one gets a set $\D_0$ dense in $\H$.

%%%%%%
\bigskip
By taking vacuum expectation values (v.e.v.) of $\star$-products of fields
one can introduce different kinds of $n$-point functions, that will be
(mere) distributions:
Wightman functions
\be
\W^{\alpha_1,...,\alpha_n}(x_1,...,x_n)=
\left(\Psi_0,\varphi^{\alpha_1}(x_1)\star...\star\varphi^{\alpha_n}(x_n) \Psi_0\right),
\ee
where $\alpha_1,...\alpha_n$ enumerate possible different field species and/or
$SL(2, \b{C})$-tensor (spinor, vector,...) components,
or (their linear combinations) Green's functions
\be
G^{\alpha_1,...,\alpha_n}(x_1,...,x_n)=
\left(\Psi_0,T\left[\varphi^{\alpha_1}(
x_1)\star...\star\varphi^{\alpha_n}(x_n)\right] \Psi_0\right),
\ee
or retarded functions, etc. In the second definition
 there appears the  time-ordering $T$,
but there is in fact no ambiguity in defining $T$ as on commutative Minkowski
space\footnote{In the standard
approach \cite{LiaSib02,BozFisGroPitPutSchWul03,AlvVaz03} this was found to
be safe and unambiguous only in the case of space-time commutativity
($\theta^{0i}= 0$), which gives commuting time variables $x^{0}_i$, so that
time-ordering commutes with the $\star$-product.},
\bea
&&T\left[\varphi^{\alpha_1}(x_1)\!\star\!\varphi^{\alpha_2}(x_2)\!\star\!...\!\star\!
\varphi^{\alpha_n}(x_n)\right]
=\varphi^{\alpha_1}(x_1)\!\star\!\varphi^{\alpha_2}(x_2)\!\star\!...\!\star\!
\varphi^{\alpha_n}(x_n)\vartheta(x_1^0\!-\!x_2^0)...\qquad\label{TimeOrdering}
\\[8pt]
&&...\vartheta(x_{n\!-\!1}^0\!-\!x_n^0)
+\varphi^{\alpha_2}(x_2)\!\star\!\varphi^{\alpha_1}(x_1)\star
\varphi^{\alpha_3}(x_3)...\varphi^{\alpha_n}(x_n)
\vartheta(x_2^0\!-\!x_1^0)...
\vartheta(x_{n\!-\!1}^0\!-\!x_n^0)+...,    \qquad   \nonumber
\eea
as this definition involves multiplication by the $\star$-central
$\vartheta(x_i^0\!-\!x_j^0)$  ($\vartheta$  denotes the Heavyside
function). [The $\star$'s preceding all $\vartheta$ can be and have
been dropped, by (\ref{startrivial}).]

Arguing as in \cite{StrWig63} for ordinary  QFT, exactly the same
properties follow from QM1-3 (alone).
Applying a pure translation, from QM2
we find that  {\bf Wightman and Green's
functions are translation invariant} and therefore may depend only on
the commutative spacetime variables  $\xi^{\mu}_i$:
\bea
\W^{\alpha_1,...,\alpha_n}(x_1,...,x_n)&=&
 W^{\alpha_1,...,\alpha_n}( \xi_1,..., \xi_{n\!-\!1}),\\
\G^{\alpha_1,...,\alpha_n}(x_1,...,x_n) &=&
 G^{\alpha_1,...,\alpha_n}( \xi_1,..., \xi_{n\!-\!1}).
\eea
% In the following for brevity we shall use a
%multivector notation $\W(x) = \W(x_1, ...,x_n), \, x = (x_1,
%...,x_n)$, $W(\xi) =W(\xi_1, ...,\xi_{n\!-\!1}), \, \xi=(\xi_1, ...,\xi_{n\!-\!1})$.
Moreover, from QM3, QM2, QM1 it respectively follows

\medskip
\vspace{1mm}\noindent  {\bf W1.} $\W^{\alpha_1,...,\alpha_n}(x_1, ...,x_n)$
are tempered distributions depending only on the $\xi_i$.

\vspace{1mm}\noindent{\bf W2.} ({\bf Spectral condition}) The
support of the Fourier transform $\tW$ of $W$ is contained in the
product of forward cones,   i.e.
\be
{ \tW^{\alpha_1,...,\alpha_n}(q_1, ...q_{n\!-\!1}) =
0,\qquad\mbox{if }\:\exists j:\quad q_j \notin \bV.}
\ee

\vspace{1mm}
\noindent{\bf W3.} ({\bf Hermiticity and Positivity})
The transformation properties of Wightman functions
under complex conjugation follow from
$$
\overline{\left(\Psi_0,\varphi^{\alpha_1}(x_1)\star...
\star\varphi^{\alpha_n}(x_n) \Psi_0\right)}=
\left(\Psi_0,\varphi^{\alpha_n\dagger}(x_1)\star...
\star\varphi^{\alpha_1\dagger}(x_n) \Psi_0\right).
$$
In particular, if all fields are Hermitean scalar then
$\overline{\W(x_1, ...,x_n)}=\W(x_n, ...,x_1)$.
 For any terminating sequence
 $\underline{f} = (f_0, f_1, ...f_N)$, $f_j \in {\cal S}(\b{R}^4)^j$ one
has \footnote{This is the transcription of positivity of the norm
of any state of the form
$$
\Psi_{\underline{f}} = f_0 \Psi_0+
\varphi(f_1)\,\Psi_0 + \varphi(f_2^{(1)})\,
\varphi(f_2^{(2)})\Psi_0 +...,
$$
where $\underline{f} = (f_0, \,f_1,... f_N)$, $\,f_j =
\prod_{k = 1}^j f_j^{(k)}(x_k)$.}
\be
{ \sum_{j,k}\int d x \,d y
\,\bar{f}_j(x_j, ...x_1)\,f_k(y_1, ...,y_k)\,
\W(x_1,...x_j; y_1,
...,y_k))\, \geq 0.}
\ee

\vspace{2mm}
We now recall the ordinary relativistic conditions on QFT:

\vspace{1mm}\noindent {\bf R1.} ({\bf Lorentz Covariance}) The
universal covering group $SL(2,\b{C})$ of the restricted Lorentz group
is a symmetry of the theory and
is represented  on $\H$ by
(strongly continuous) unitary operators $U(A)$. The fields transform
covariantly under the inhomogeneous $SL(2,\b{C})$
(i.e. generalized Poincar\'e) transformations $U(a, \,A)
= U(a)\,U(A)$:
\be
U(a, A) \,\varphi^\alpha(x)\,U(a, A)^{-1} =
S^\alpha_{\beta}(A^{-1})\,\varphi^\beta\big(\Lambda(A) x + a\big),
\label{transf} \ee
with $S$ a finite dimensional representation of $SL(2,\b{C})$
and $ \Lambda(A)$ the Lorentz transformation associated to $A\in  SL(2,\b{C})$.

\vspace{1mm}\noindent {\bf R2.} ({\bf Microcausality or locality})
The fields either commute or anticommute at spacelike separated
points
\be
{ [ \, \varphi^\alpha(x), %    \stackrel{\star},
\,\varphi^\beta(y)\,]_{\mp} = 0,
\qquad\mbox{for}\,\,\,(x - y )^2 < 0.}             \label{fieldcomrel}
\ee

\medskip
As a consequence of QM2,R1  in QFT on commutative Minkowski
space one finds

\vspace{1mm}\noindent {\bf W4.} ({\bf Lorentz Covariance of Wightman
functions})
\be
 \W^{\alpha_1...\alpha_n} \big(\Lambda(A)x_1, ...,
\Lambda(A)x_n\big)
 =S^{\alpha_1}_{\beta_1}(A)...
 S^{\alpha_n}_{\beta_n}(A)
 \W^{\beta_1...\beta_n}(x_1, ...,x_n).\quad
\label{LorCov} \ee
%  \be
%\W(x_1, ...x_n)  \equiv W(\xi_1, ...\xi_{n-1}) =
%\W(\Lambda x_1, ...,\Lambda x_n)  \equiv W(\Lambda \xi_1, ...,\Lambda \xi_{n-1})
%.                \label{LorCov}
%\ee
Wightman functions
of scalar fields are  Lorentz invariant.
(Similarly for Green functions).

\medskip
In order to translate R1 into a corresponding condition R1$_\star$ in
the twisted Hopf algebra setting we could go either to the infinitesimal
formulation (i.e. first to $\cal P$, and then deform to $H$), or to the dual
functions-on-the-group Hopf algebra. We do not attempt this here, because it
would be rather technical (especially translating the strong continuity
requirement), and moreover some subtlety might be hidden in the interplay of
active (or system) and passive (or coordinates) twisted Poincar\'e
transformations appearing at the two sides of (\ref{transf}). We content
ourselves with requiring the deformed analog of W4,
which should  follow from R1$_\star$ however this will look like, namely
that Wightman (and Green) functions transform under a twisted version of
(\ref{LorCov}), in particular are invariant if all involved fields are scalar.
On the other hand, as these functions should be built only in terms of the
$\xi^{\mu}_i$ and of ordinary $SL(2,\b{C})$ tensors, like
$\partial_{x^{\mu}_i}$, the isotropic tensor $\delta_\nu^\mu$, spinors, 
$\gamma$-matrices, etc.,
which are all annihilated by the action of $P_\mu$,
the action of the twist ``legs'' $\F^{(1)},\F^{(2)}$ should be trivial and
the transformation properties under the Lorentz generators should remain
undeformed: so {\it these functions should  admit exactly the same decomposition
in Lorentz tensors as in the undeformed case (in particular should be
invariant if all fields are scalar fields)}. Therefore, deferring the
formulation of R1$_\star$ to possible future works, here  {\bf  we shall
require W4 also in the deformed case} as a temporary substitute of R1.

As for R2, it is natural to ask whether in the
deformed theory one can adopt the twisted version

\vspace{1mm}\noindent {\bf R2$_\star$.} ({\bf Microcausality or locality})
The fields either $\star$-commute or $\star$-anticommute at spacelike separated
points\footnote{As already noted, space-like separation is well-defined, so
that  the latter condition makes sense.}
\be
{ [ \, \varphi^\alpha(x)   \stackrel{\star},
\,\varphi^\beta(y)\,]_{\mp} = 0,
\qquad\mbox{for}\,\,\,(x - y )^2 < 0.}             \label{fieldstarcomrel}
\ee
 and whether there also
viable alternatives. That the conditions QM1-3, W4,  R2 are independent and
compatible can be proved  arguing along the lines \cite{StrWig63}; in
particular compatibility is proved by showing that they can be fulfilled by
free fields (see next section). We thus find in particular that  {\bf
the noncommutativity structure of a
Moyal-Weyl space is compatible with locality R"$_\star$}!
Whether reasonable weakenings of R2$_\star$ exist is in fact an open question
also in the ordinary theory.
Phenomenology suggests that rhs(\ref{fieldcomrel}) should at least rapidly decrease
with increasing space-like distances, if it is not zero. On the other hand, the
same results as in  \cite{WigPetVla,BorPoh68} should apply, namely
requiring that rhs(\ref{fieldcomrel}) is zero only in some space-like separated
open subsets  (see \cite{WigPetVla}, or Theorem 4-1 in \cite{StrWig63}), or
is a $c$-number  decreasing faster than an exponential with
space-like distances  \cite{BorPoh68},
are actually only apparent weakenings, in that they imply again R2.

\medskip
As consequences of R2$_\star$ one again finds \cite{StrWig63}

\vspace{1mm}\noindent {\bf W5.} ({\bf Local commutativity conditions})
If $(x_j \!-\! x_{j\!+\!1})^2 \!<\! 0$ then 
\be
 \W^{\alpha_1...\alpha_n} (x_1, ...
x_j, x_{j\!+\!1}, ...x_n) = \pm 
\W^{\alpha_1...\alpha_{j\!+\!1}\alpha_j...\alpha_n} 
(x_1, ...x_{j\!+\!1}, x_j,
...x_n);
\ee
the sign is negative if 
$\varphi^{\alpha_j},\varphi^{\alpha_{j\!+\!1}}$ 
$\star$-anticommute, is positive otherwise.

\vspace{1mm}\noindent {\bf W6}. ({\bf Cluster
property}) For any spacelike vector $a$ and for $\lambda \to \infty$
\be
 \W^{\alpha_1...\alpha_n}\!(x_1,\! ...x_j, x_{j\!+\!1}
\! +\! \lambda a,\! ...,x_n \!+\! \lambda
a) \to \W^{\alpha_1...\alpha_j}\!(x_1,\! ...,x_j)\,
\W^{\alpha_{j\!+\!1}...\alpha_n}\!(x_{j\!+\!1},\! ...,x_n),
\ee
(convergence is in the distributional sense); this is true
also with permuted coordinates.

In the proof of W6 the uniqueness of the invariant state $\Psi_0$ plays
an essential role.

\bigskip
Summarizing, we end up with a QFT framework on these NC spaces with  
QM1-3, W4,  R2$_\star$ or alternatively with exactly the same constraints 
 W1-6 on Wightman  functions as in ordinary QFT on
Minkowski space. Reasoning as described in \cite{StrWig63,Jost,Haa96}, one
should be able to prove the same, other well-known fundamental results in
ordinary QFT:

\begin{enumerate}

\item That Wightman functions are boundary values
$$
W^{\alpha_1...\alpha_n}(\xi_1, ...\xi_{n-1})=\lim\limits_{\eta_1,...,\eta_{n\!-\!1}\to 0} W^{\alpha_1...\alpha_n}(\zeta_1,
...\zeta_{n-1})
$$
of holomorphic functions
$W(\zeta_1, ...\zeta_{n-1})$
in the complex variables $\zeta_i=\xi_i-i \eta_i$, the domain of holomorphy
being $\{\zeta_1, ...\zeta_{n-1}\, |\, \eta_j\in V^+\}$.

\item The analogs of the Spin-Statistics and CPT theorems.

\item That the cluster property  W6 implies (Haag-Ruelle theory)
 the existence of asymptotic (free) fields and, under the assumption of
asymptotic completeness ($\H=\H^{in}=\H^{out}$), of a unitary $S$-matrix.
This allows to derive \cite{Hepp} the LSZ formulation of QFT, and
subsequently dispersion relations for scattering amplitudes, etc.

\item That the Wightman
functions have an analytic continuation to the socalled Euclidean
points, thus allowing to derive the existence and the
general properties of Euclidean QFT with the analog of Schwinger functions.

\end{enumerate}

We stress that these results should hold for all $\theta^{\mu\nu}$,
and not only if $\theta^{0i}=0$ as in the approach e.g. of
\cite{ChaPreTur05,ChaMnaNisTurVer06}.

\section{Free fields}
\label{Free}

As in ordinary QFT, things become much more definite
for {\it free} fields. By (\ref{derivatives}), the kinetic differential
operators (D'Alambertian, Dirac operator, etc) remain undeformed, therefore so
will remain the free field equations and the  consequent constraints on
Wightman, Green's functions and on the field commutation relations.
For simplicity we stick to the case of a  free Hermitean scalar field
$\varphi_0(x)$ of mass $m$:
\be
(\Box_{x}+m^2)\varphi_0=0.                             \label{freefield}
\ee
In momentum space this becomes
$(p^2-m^2)\tilde \varphi(p)=0$, so the spectrum is contained in (the two
sheets of) the hyperboloyd $p^2=m^2$. We can therefore Fourier decompose
$\varphi_0(x)$ into a positive and a negative frequency part in a
(twisted) Lorentz invariant way,
\be
\ba{l}
\varphi_0(x)= \varphi_0^+(x)+\varphi_0^-(x)\\[8pt]
\varphi_0^+(x)\!:=\!\int \!d\mu(p)\,
e^{-ip\cdot x}a^p ,\\[8pt]
\varphi_0^-(x)\!:=\!\int \!d\mu(p)\,
a_p^\dagger  e^{ip\cdot x} \!=\!\big(\varphi^+_0(x)\big)^\dagger,
\ea                                               \label{fielddeco}
\ee
where $d\mu(p)=\delta(p^2\!-\!m^2)\vartheta(p^0)d^4p=
dp^0\delta(p^0\!-\!\omega_{\bf p})d^3{\bf p}/2\omega_{\bf p}$
is the invariant measure ($\omega_{\bf p}\!:=\!\sqrt{{\bf p}^2+m^2}$).
From (\ref{freefield}) it immediately follows
$(\Box_\xi+m^2)W(\xi)=0$ or equivalently $(p^2-m^2)\tilde W(p)=0$
in momentum space,
%(at least for $x\neq y$ (?)),
whence the Fourier decomposition
$$
W(x\!-\!y)=\int \!d\mu(p)\,[w^+(p)e^{-ip\cdot(x\!-\!y)}+w^-(p)e^{ip\cdot
(x\!-\!y)}].
$$
On the other hand, using QM1-3 one finds first $\varphi^+_0(x)
\Psi_0=0$, i.e. $a^p\Psi_0=0$, then
$$
W(x\!-\!y)= \left(\Psi_0,\varphi_0(x)\star\varphi_0(y)
\Psi_0\right)=\left(\Psi_0,\varphi^+_0( x)\star\varphi^-_0(y) \Psi_0\right),
$$
showing that $x$ (resp. $y$) is associated only to the positive (resp. negative)
frequencies, i.e. $w^-(p)$ has to vanish, and $w^+(p)$ has to be positive.
But by W4  $w^+(p)$ has to be Lorentz invariant, i.e. constant, so we
conclude that up to a positive factor $W$ is given by
\be
W(x-y)= -i F^+(x-y), \qquad \qquad     F^+(\xi):=i
\int \frac{d\mu(p)}{(2\pi)^3}e^{-ip\cdot \xi}        \label{2W}
\ee
and therefore coincides with the undeformed counterpart. Moreover,
\be
\big(\Psi_0,a^pa_q^\dagger \Psi_0\big)=2\omega_{\bf p}\delta^3({\bf
p}\!-\!{\bf q})                                         \label{vev}
\ee
as in the undeformed case. The 2-point Green's function is now immediately
obtained as the time-ordered combination of $W(x\!-\!y)$ and $W(y\!-\!x)$:
\be
\ba{l}
\G(x,y):=\left(\Psi_0,T\left[\varphi_0(x)\star\varphi_0(y)\right]\Psi_0\right)=  G(
x\!-\!y),\\[8pt]
 G(\xi):=-i \left[\vartheta(\xi^0) \, F^+(\xi) +\vartheta(-\xi^0)\,
F^+(-\xi)\right]= \int \frac{d^4p}{(2\pi)^4} \frac{e^{-ip\cdot
\xi}}{p^2-m^2+i\epsilon},
\ea                      \label{propagator}
\ee
and therefore coincides with (the undeformed) Feynman's propagator.
Note that (\ref{2W}), (\ref{vev}), (\ref{propagator}) {\it are independent of
{\bf R2$_\star$} or any other assumption about the field commutation relations}, which
we have not used in the proof.

\medskip
On the other hand, if one postulates all the axioms of
the preceding section (including  {\bf R2$_\star$}) and reasons as in the proof of the
Jost-Schroer theorem, Thm 4-15\footnote{More precisely, as it is done after the
proof that (\ref{freefield}) follows from (\ref{2W}).} in
\cite{StrWig63},  one  proves up to a positive factor the {\bf free field
commutation relation}
\be
[\varphi_0(x)\stackrel{\star},\varphi_0(y)]=
i\,F(x-y), \qquad\qquad
F(\xi):=F^+(-\xi)-F^+(\xi),                        \label{freecomm}
\ee
{\bf which coincides with the undeformed one}. Incidentally, this can be proved
also from just the free field equation and the assumption that the commutator
is a (twisted, and therefore also untwisted) Poincar\'e invariant $c$-number
(see e.g. \cite{Schweber}, page 178-179). Applying $\partial_{y^0}$ to (\ref{freecomm}) and then setting $y^0=x^0$ [as already noted, this is compatible
with (\ref{summary})] one finds {\bf the canonical commutation relation}
\be
[\varphi_0(x^0,{\bf x})\stackrel{\star},\dot\varphi_0(x^0,{\bf y})]=
i\,\delta^3({\bf x}-{\bf y}).                      \label{Cancomm}
\ee

As a consequence of (\ref{freecomm}), also the  $n$-point Wightman
functions coincide with the undeformed ones, i.e. vanish if $n$ is odd  and are
sum of  products of two point functions (this is the so-called factorization)
if $n$ is even. This of course agrees with the cluster property W6.

\medskip
Free fields fulfilling (\ref{freecomm}) can be obtained from
(\ref{fielddeco}) plugging creation, annihilation operators
fulfilling commutation relations deformed  so as to
compensate the spacetime noncommutativity. The first 
possibility\footnote{In this and
other proofs one has to use the following properties of  exponentials.
Recalling the Baker-Campbell-Hausdorff formula $e^Ae^B=e^{A+B+C}$ (with
$C:=[A,B]/2$ commuting with $A,B$) one finds
%$$
%e^{ip\cdot x}e^{iq\cdot y}=e^{ip\cdot x+iq\cdot y-\frac i2 p\theta q}
%\qquad\Rightarrow \qquad e^{ip\cdot x}e^{-ip\cdot y}=e^{ip\cdot (x- y)},
%$$
%as $p\theta p=0$. These relations hold in particular for $y=x$.
%More generally, by iteration of the previous result one finds
%\be
%e^{ip_1\cdot x_1}e^{ip_2\cdot x_2}...e^{ip_n\cdot x_n}
%=e^{i\sum\limits_{j=1}^n p_j\cdot x_j-\frac i2 \sum\limits_{j<k}p_j\theta p_k},
%\ee
%which holds also if $x_j=x_k$ for some $j,k$.
$$
e^{ip\cdot x}\!\star\! e^{iq\cdot  y}=e^{ip\cdot  x\!+\!iq\cdot  y\!-\!\frac i2 p\theta q}
\quad\qquad\Rightarrow \quad\qquad e^{ip\cdot x}\!\star\! e^{iq\cdot  y}=
e^{iq\cdot  y}\!\star\! e^{ip\cdot x}e^{-i p\theta q}, \qquad
e^{ip\cdot  x}\!\star\! e^{-ip\cdot  y}=e^{ip\cdot
(x\!-\!y)};
$$
the last follows from the first and $p\theta p=0$. These relations hold in
particular for $y=x$. More generally, by iteration of the previous result
one finds
\be
e^{ip_1\cdot  x_1}\!\star\! e^{ip_2\cdot  x_2}\!\star ...\star\! e^{ip_n\cdot  x_n}
=\exp\left(i\sum\limits_{j=1}^n p_j\cdot  x_j\!-\!\frac i2
\sum\limits_{j<k}p_j\theta p_k\right),
\ee
which holds also if $x_j=x_k$ for some $j,k$.} 
is to require 
\be
\ba{l}
a^{\dagger}_p  a^{\dagger}_q= e^{i
 p\theta' q}\,     a^{\dagger}_q  a^{\dagger}_p, \\[8pt]
a^p  a^q=
e^{i p\theta' q} \,    a^q  a^p,  \\[8pt]
a^p a^{\dagger}_q=
e^{-i  p\theta' q} \,  a^{\dagger}_q  a^p+2
\omega_{\bf p}\delta^3({\bf p}\!-\!{\bf q}), \\[8pt]
[a^p,f(x)]=[a^{\dagger}_p,f(x)]=0,
\ea \qquad\qquad\theta'=\theta                            \label{aa+cr}
\ee
(where $p\theta q:=p_{\mu}\theta^{\mu\nu}q_{\nu}$
for any 4-vectors $p,q$), as adopted e.g. in
\cite{BalPinQur05,BalGovManPinQurVai06,LizVaiVit06,Abe06}
[see also the bibliographical notes after (\ref{aa+cr"})]. 
In the sequel we wish to
consider and compare also other choices of the  parameters $\theta'{}^{\mu\nu}$.
The choice $\theta'=0$ gives the CCCr (canonical commutation relations),  assumed
in most of the literature,  explicitly \cite{DopFreRob95}
or implicitly, either in operator  (e.g. apparently in \cite{ChaNisTur06,
ChaPreTur05,ChaMnaNisTurVer06,Tur06}), or in
path-integral approach to quantization (see e.g. \cite{Fil96},
\cite{DouNek01,Sza03} and most references therein).   Note that the last term
in the third equation is fixed by (\ref{vev}). Correspondingly, one finds the
field $\star$-commutation relations
\be
\varphi_0(x)\star\varphi_0(y)=e^{i  \partial_x(\theta-\theta') \partial_y}
\varphi_0(x)\star\varphi_0(y)+i\,F(x-y),             \label{nonlocalcom}
\ee
which are non-local unless $\theta'=\theta$.
As said, the authors of
\cite{BalGovManPinQurVai06,BalPinQur05,LizVaiVit06,Abe06} 
adopt $\theta'=\theta$. In \cite{LizVaiVit06}
commutation relations of the form (\ref{freecomm}) are proposed in a 1+1 dimensional model
in order to close the chiral current algebra; in \cite{Abe06} (\ref{freecomm}) are proposed
in any dimension, although only for scalar fields and for $\theta^{0i}=0$; whereas the authors of \cite{BalPinQur05,BalGovManPinQurVai06} find non-local
relations [see formula (3.23) in \cite{BalGovManPinQurVai06}] similar to (\ref{nonlocalcom}), because
they do not perform the $\star$-product between functions of different sets
$x,y$ of coordinates.

Let us consider two typical contributions to the 4-point Wightman function:
$$
\W(x_1,x_2,x_3,x_4)=W(x_1\!-\!x_2)W(x_3\!-\!x_4)+
e^{i  \partial_{x_2}(\theta-\theta') \partial_{x_3}}
W(x_1\!-\!x_3)W(x_2\!-\!x_4)+...
$$
The first term at the rhs comes from the v.e.v.'s
of $\varphi_0(x_1)\!\star\!\varphi_0(x_2)$ and
$\varphi_0(x_3)\!\star\!\varphi_0(x_4)$; it is Lorentz invariant by
(\ref{2W}) and factorized. The second, nonlocal term comes from  the v.e.v.'s
of $\varphi_0(x_1)\!\star\!\varphi_0(x_3)$ and of
$\varphi_0(x_2)\!\star\!\varphi_0(x_4)$,  after using (\ref{nonlocalcom}) to
commute $\varphi_0(x_2),\varphi_0(x_3)$.
Only if $\theta'=\theta$ it is Lorentz invariant and factorizes
into $W(x_1\!-\!x_3)W(x_2\!-\!x_4)$. As it depends only on $x_1\!-\!x_3$,
$x_2\!-\!x_4$, it is invariant under the replacements
$(x_1,x_2,x_3,x_4)\to(x_1,x_2\!+\!\lambda a,x_3,x_4\!+\!\lambda a)$,
even in the limit $\lambda\to\infty$
By taking $a$ space-like, we conclude that if $\theta'\neq\theta$ $\W$
violates W4 and W6, as expected.

\medskip
We present a second way to realize
(\ref{freecomm}), which at first sight might appear ``exotic'', but we are
going to theoretically motivate elsewhere. It follows
from assuming nontrivial transformation laws
 $P_\mu\trc a^{\dagger}_p=p_\mu a^{\dagger}_p$,
$P_\mu\trc a^p=-p_\mu a^p$ and extending the $\star$-product law
(\ref{starprod}) also to $a^p,a^{\dagger}_p$. It amounts to
choosing $\theta'=-\theta$ in (\ref{aa+cr}) [inserting for uniformity of notation
a  $\star$-symbol in each product, also in (\ref{fielddeco})] and
to adding nontrivial commutation relations between the $a^p,a^{\dagger}_p$ and
functions, in particular exponentials, of the form
\be
\ba{l}
a^{\dagger}_p \star a^{\dagger}_q= e^{-i
 p\theta q}\,     a^{\dagger}_q \star a^{\dagger}_p, \\[8pt]
a^p\star  a^q=
e^{-i p\theta q} \,    a^q \star  a^p, \\[8pt]
a^p\star  a^{\dagger}_q=e^{i p\theta q} \,  a^{\dagger}_q \star a^p+
2\omega_{\bf p}\delta^3({\bf p}\!-\!{\bf q}),\\[8pt]
a^p\star e^{iq\cdot x} =e^{-ip\theta q} \,e^{iq\cdot x}\star a^p,
\qquad\qquad a^{\dagger}_p\star e^{iq\cdot x} =
e^{ip\theta q} \,e^{iq\cdot x}\star a^{\dagger}_p.
\ea                           \label{aa+cr'}
\ee
As a consequence, $[\varphi_0(x)\stackrel{\star},f(y)]=0$.
In other words, the first three relations in (\ref{aa+cr'}) define an
example of a general deformed Heisenberg algebra \cite{Fio95}
\be
\ba{l}
a^q \star a^p= R^{qp}_{rs} \: a^s\star  a^r                   \\[8pt]
a^{\dagger}_p \star a^{\dagger}_q=  R^{sr}_{pq} \: a^{\dagger}_r
\star a^{\dagger}_s  \\[8pt]
a^p \star a^{\dagger}_q=\delta^p_q+ R^{rp}_{qs} \: a^{\dagger}_r
\star a^s
\ea              \label{aa+cr"}
\ee
covariant under a triangular Hopf algebra $H$.  Here the $R$-matrix is
the universal $\R$ in the infinite-dimensional representation of $H$ spanned by
the basis of vectors $a^{\dagger}_p$, the (on-shell)
4-momenta $p,q,r,s$ playing the role of (continuous) indices, and
$\delta^p_q\!=\!2\omega_{\bf p}\delta^3({\bf p}\!-\!{\bf q})$ is Dirac's delta
(up to normalization).  [The first three relations in (\ref{aa+cr}) also can be
considered of the form (\ref{aa+cr"}) after a replacement $\theta\!\to
\!-\theta'$]. Such $a^p,a^{\dagger}_p$ can be realized \cite{Fio95}
as composite of operators $c^p,c^{\dagger}_p$ fulfilling the CCR 
(for the case at hand of the $\theta$-deformed Poincar\'e this
has been done also in \cite{BalManPinVai05}), so that they act on the same
(undeformed) Fock space. In doing so one finds that the action of
$P_\mu$ can be realized as a commutator with the operator
$\tilde P_\mu=\int d\mu(p)c^\dagger_p c^p =\int d\mu(p)a^\dagger_p a^p$.

As historical remarks we add that, up to normalization of $R$, and
with $p,q,r,s\in\{1,...,N\}$,  relations (\ref{aa+cr"}) are also identical to
the ones defining the older $q$-deformed Heisenberg algebras of
\cite{PusWor89,WesZum90}, based on a quasitriangular $\R$ in (only) the
{\it fundamental} representation of $H=U_qsu(N)$;
allowing a different (possibly infinite-dimensional) representation 
has been considered in
\cite{GroMadSte} for the $U_qsu(2)$-covariant
quantization of fields on the $q$-deformed fuzzy sphere.
Going further back in the past, (\ref{aa+cr'}-\ref{aa+cr"}) are 
reminiscent of the Zamoldchikov-Faddeev \cite{ZamZam79,Fad80}
algebra, generated by deformed creation/annihilation operators of scattering
states of some completeley integrable  1+1-dimensional QFT; there
again the indices are discrete, but the $R$-matrix depends 
on a (continuous) spectral parameter, the rapidity of the particles. 
In \cite{Kul81} the Zamoldchikov-Faddeev creation/annihilation
operators have been realized as acting
on the (undeformed) Fock space.

\section{Theoretical developments. Perturbative expansion for interacting
QFT}
\label{Interact}

Normal ordering should be
a $\A_{\theta}^n$-bilinear map of the field algebra into itself, such that
any normal ordered expression has a trivial v.e.v., in particular
$:\!\1\!: \:=0$.
Applying it to the sides of (\ref{aa+cr}) we find that it is consistent to
define
\be
:\!a^p a^q\!: \:=a^p a^q, \qquad :\!a^{\dagger}_p a^q\!: \:=a^{\dagger}_p a^q,
\qquad :\!a^{\dagger}_p a^{\dagger}_q\!: \:= a^{\dagger}_p a^{\dagger}_q,
\qquad :\! a^p  a^{\dagger}_q\!: \: \:=\:\,
a^{\dagger}_q  a^p e^{-i  p\theta' q}.
\ee
The phase\footnote{The authors of \cite{BalPinQur05}
omit this phase. However, their conclusions
about the $S$-matrix being undeformed remain valid, as the
effect of this phase disappears when exploiting global energy-momentum
conservation.} in the last relation is to account for (\ref{aa+cr})$_3$ and
$:\!\1\!: \:=0$. More generally, it is consistent to {\bf define normal
ordering} on any monomial in $a^p,a^{\dagger}_q$ as a map which 
reorders all  $a^p$ to the right of all $a_q^{\dagger}$ introducing a factor
$e^{-i  p\theta' q}$ for each flip $a^p \leftrightarrow a_q^{\dagger}$.
For $\theta'=0$ one finds the undeformed definition. Using
$\A_{\theta}^n$-bilinearity normal ordering is extended to fields.

\medskip
We first consider the assumptions leading to (\ref{freecomm}), namely
(\ref{aa+cr}) or (\ref{aa+cr'}). One finds that exactly as in the
undeformed case it maps each monomial $M$
in the fields (and their derivatives) into itself minus all lower degree
monomials  obtained by taking v.e.v.'s  of pairs of fields
appearing in $M$, e.g.
\be
\ba{l}
:\!\varphi_0(x) \!: \:  \: \:=\,
\varphi_0(x) \\[8pt]
:\!\varphi_0(x) \star \varphi_0(y)\!: \:  \: \:=\,
\varphi_0(x) \star \varphi_0(y) -\left(\Psi_0,\varphi_0(x) \star
\varphi_0(y)\Psi_0\right)\\[8pt]
:\!\varphi_0(x) \!\star\! \varphi_0(y)\!\star\! \varphi_0(z)\!: \:  \: \:=\,
\varphi_0(x) \!\star\! \varphi_0(y)\!\star\! \varphi_0(z) -\left(\Psi_0,\varphi_0(x)
\!\star\! \varphi_0(y)\Psi_0\right)\varphi_0(z)\\[8pt]
\qquad\qquad\qquad -\left(\Psi_0,\varphi_0(x)
\!\star\! \varphi_0(z)\Psi_0\right)\varphi_0(y)
- \varphi_0(x)\left(\Psi_0,\varphi_0(y)
\!\star\! \varphi_0(z)\Psi_0\right)\\[8pt]
\qquad\qquad\qquad\qquad\qquad\qquad...
\ea                                  \label{NorOrdering}
\ee
By construction $\left(\Psi_0,:\!M\!:\Psi_0\right)=0$.
These are well-defined operators also in the limit of coinciding
coordinates (e.g. $y\to x$).
The above substractions amount to
flipping step by step each $\varphi_0^+(x)$ to the right of each
$\varphi_0^-(y)$. For instance
on $\varphi_0^{\varepsilon}(x)\varphi_0^{\eta}(y)$
($\varepsilon,\eta\in\!\{-,+\}$) normal ordering acts as the identity
unless $\varepsilon\!=\!+$ and $\eta\!=\!-$, whereas
$$
%&&:\!\varphi_0^+(x) \star \varphi_0^+(y)\!:  \: \:=\,
%\varphi_0^+(x) \star \varphi_0^+(y) \qquad\quad
%:\!\varphi_0^-(x) \star \varphi_0^-(y)\!: \ea   \label{propagator}\:\:=\, \varphi_0^-(x) \star
%\varphi_0^-(y)\nn
%&& :\!\varphi_0^-(x) \star \varphi_0^+(y)\!: \:\:=\, \varphi_0^-(x) \star \varphi_0^+(y)
%\qquad\quad
:\!\varphi_0^+(x) \star \varphi_0^-(y)\!: \:\:=\, \varphi_0^-(y) \star \varphi_0^+(x).
$$
As a consequence we find that for any monomial
$M'$ obtained from $M$ by permutation of the field factors
$:\!M\!:\: =\: :\!M'\!:$, for instance
\be
\ba{c}
:\!\varphi_0(x) \star \varphi_0(y)\!: \:  \: \:=\:\,
:\!\varphi_0(y) \star \varphi_0(x)\!: \\
...
\ea
\label{reordering}
\ee
Moreover, as $\varphi_0(x)$ is Hermitean, any normal ordered monomial
$:\!\varphi_0(x_1) \!\star\!... \!\star\!\varphi_0(x_n)\!:$
is ({\it a fortiori} for coinciding coordinates). Summing up, under these
assumptions normal ordering (\ref{NorOrdering}) and its properties  are written
in terms of the fields exactly as in the undeformed setting
(apart from the occurrence of the $\star$-product symbol).
Since the same occurs with time-ordering (\ref{TimeOrdering}),
another straightforward consequence is that {\bf the same   Wick theorem
will hold}:
$$
\ba{l}
T\big[\varphi_0(x) \star \varphi_0(y)\big]=\: :\!\varphi_0(x) \star
\varphi_0(y)\!: +\Big(\Psi_0,T\big[\varphi_0(x) \star \varphi_0(y)\big]\Psi_0\Big) \\[8pt]
T\big[\varphi_0(x) \!\star\! \varphi_0(y)\!\star\! \varphi_0(z)\big]
 =\, :\!\varphi_0(x) \!\star\! \varphi_0(y)\!\star\! \varphi_0(z)\!:
 +\Big(\!\Psi_0,T\big[\varphi_0(x) \star \varphi_0(y)\big]\Psi_0\!\Big)
\, :\!\varphi_0(z)\!: \\[8pt]
\qquad\qquad+\Big(\!\Psi_0,T\big[\varphi_0(x) \star \varphi_0(z)\big]\Psi_0\!\Big)
\, :\!\varphi_0(y)\!:+\Big(\!\Psi_0,T\big[\varphi_0(y) \star \varphi_0(z)\big]
\Psi_0\!\Big)\, :\!\varphi_0(z)\!:\\[8pt]
\qquad\qquad\qquad\qquad\qquad\qquad...
\ea
%\label{wick}
$$

\bigskip
Let us apply now 
{\bf time-orderd perturbation theory} to an interacting field.
We use the Gell-Mann--Low   formula (rigorously valid under the assumption of
asymptotic completeness, $\H=\H^{in}=\H^{out}$)
\be
 G( x_1,...,x_n)  =\frac{
\left(\Psi_0,T\left\{\varphi_0(x_1)\star ...\star \varphi_0(x_n)\star
\exp\left[ -i\lambda \int dy^0  \ H_I(y^0)\,\right]\right\}
\Psi_0\right)} {\left(\Psi_0,T\exp\left[ -i \int dy^0
\ H_I(y^0)\,\right]\Psi_0\right)}.
\label{formal}
\ee
Here $\varphi_0$ denotes the free ``in'' field, i.e. the incoming field in the
interaction representation, and $H_I(x^0)$ is
the interaction Hamiltonian in the interaction representation.
The derivation of (\ref{formal}) is heuristic and goes
as on commutative space. It involves  unitary evolution operators of the form 
$$
U(x^0,y^0)=\lim\limits_{N\to\infty}
\prod\limits_{m=0}^{\stackrel{\longleftarrow}{N\!-\!1}}
\exp\left[-\frac i{\hbar}\frac {\Delta}N H_I(y^0\!+\!\Delta m/N)\right]=
T\exp\left[-\frac i{\hbar}\int^{x^0}_{y^0}dt H_I(t)\right],
$$ 
where $\Delta\!=\! x^0\!-\!y^0\!>\!0$  and
again $T$ always uses $\star$-central time coordinate differences as
arguments of the Heavyside function.
For the sake of definiteness  we choose
\be
H_I(x^0)= \lambda \int\! d^3 x\ :\varphi^{\star m}_0(x)\!: \,\star, \qquad\qquad
\varphi^{\star m}_0(x) \equiv\underbrace{\varphi_0(x) \star
... \star\varphi_0(x)}_{m\mbox{ times}}.           \label{HI}
\ee
This is a well-defined, Hermitean [by (\ref{reordering})] operator,
with zero v.e.v.
Expanding the exponential the generic term of order $O(\lambda^h)$ in the
numerator of (\ref{formal}) will be the v.e.v.
\be
\frac{\lambda^h}{i^h} \! \int \!d^4y_1... d^4 y_h
\Big(\Psi_0,T\left[\varphi_0(x_1)\star ...\star \varphi_0(x_n)\star
:\!\varphi^{\star m}_0(y_1)\!:  \star ...  \star
:\!\varphi^{\star m}_0(y_h)\!:  \right] \Psi_0\Big),
\label{expansion}
\ee
where we have also used the property (\ref{starIntCR}) that integration
over any space-time variable commutes with taking $\star$-products.
We proceed to evaluate this expression as in the undeformed case:
applying Wick theorem to the field monomial and the fact that
all normal-ordered field monomials  have trivial
v.e.v. we end up with exactly the {\it same} sum of
terms given by integrals over
$y$-variables, as in the undeformed case, of products of free propagators
(\ref{propagator}) having coordinate differences as arguments. Each of these
terms is represented by a Feynman diagram. So the result for the generic term
(\ref{expansion}) will be the {\it same} as the undeformed one. On the other
hand, the generic term of order $O(\lambda^h)$ in the  denominator of
(\ref{formal}) will be a special case of (\ref{expansion}), the one with
$n=0$. Summing up, numerator and denominator of (\ref{formal}), and therefore
also the {\bf Green functions (\ref{formal}) coincide with the undeformed ones}
(at least perturbatively). They can be computed by Feynman diagrams with the
undeformed Feynman rules.

In other words, at least  perturbatively, {\bf
this QFT is completely equivalent to the
undeformed counterpart}, and therefore also pathologies like UV-IR mixing
disappear. Thus, also for the interacting theory the $a_p,a^\dagger_p$
and the spacetime noncommutativities somehow compensate each other.

\bigskip
We now sketch  how perturbation theory changes
if $\theta'\neq \theta$, starting from normal ordered field monomials.
Relations (\ref{aa+cr}) lead to a non-local (pseudodifferential) operator for
each flip of a $\varphi_0^+(x)$ to the right of a $\varphi_0^-(y)$, e.g.
$$
:\!\varphi_0^+(x) \star \varphi_0^-(y)\!: \:\:=\,
e^{i \partial_x(\theta\!-\!\theta') \partial_y}\varphi_0^-(y) \star
\varphi_0^+(x), $$
whereas on $\varphi_0^{\varepsilon}(x)\varphi_0^{\eta}(y)$
with $(\varepsilon,\eta)\!\neq\!(+,-)$ normal ordering still acts as the identity.
As a consequence, property (\ref{reordering}) and Wick theorem are modified,
so are the Feynman rules, and UV/IR mixing for nonplanar Feynman
diagrams reappears. Just to get a feeling one can consider the
$\lambda\varphi^{\star 4}$ theory without normal ordering and, as in
\cite{MinRaaSei00}, one finds UV/IR mixing already in several contributions
(of nonplanar tadpole diagram type) to the
$O(\lambda)$ correction to the propagator.

\section{Final remarks and conclusions}
\label{outlook}

There is still no convincing and generally accepted formulation of
QFT on noncommutative spaces, even on the simplest
one, the Moyal-Weyl space.
One crucial aspect under debate is the form of its covariance under
space(time) symmetry transformations. In this work we have argued that a
Moyal-Weyl deformation of Minkowski space  is compatible with the
Wightman axioms (including locality) and
time-ordered perturbation theory, provided one replaces products of fields by
$\star$-products (also at different spacetime points) and the  Lorentz
covariance axiom (R1) by the appropriate twisted  version R1$_\star$ (which 
we have not formulated yet). Both for free and interacting fields the resulting QFT's appear physically equivalent to the undeformed counterparts on commutative
Minkowski space, in that their Wightman and Green's functions
coincide. This can be understood as a sort of compensation
of the effects of the $a_p,a^\dagger_p$
and of the spacetime noncommutativities, if these are matched to each other.
 (To keep the size of this work contained we have not
developed other important aspects, like those mentioned at 
the end of section \ref{QFT},
which we hope to treat soon elsewhere. For the moment, regarding
the question whether QFT on noncommutative spaces violate standard
Bose or Fermi statistics, as claimed e.g. in
\cite{BalManPinVai05,BalGovManPinQurVai06,ChaGanHazSch06},  we content with
drawing the reader's attention on Ref.'s \cite{FioSch96,Fiopro96}.)

The main positive aspect of this outcome is a way
to avoid all the additional complications (non-unitarity,
macroscopic violation of causality, UV-IR mixing
and subsequent non-renormalizability, change of statistics,...) appeared in
other approaches
and to end up with a theoretically and phenomenologically satisfactory QFT,
the undeformed theory (to the extent the latter can be considered satisfactory).
For  free field this is
achieved by matching the commutation relations of the deformed
creation/annihilation operators to the spacetime noncommutativity (however, we
have found even two different ways to realize such a matching).

The related, obvious  disappointing aspect is that in this  resulting QFT
there appears neither new physics nor a more
satisfactory formulation of the old one (e.g. by an inthrinsic UV
regularization), in that all effects of spacetime noncommutativity are confined
in an ``unobservable common noncommutative translation of all reference
frames''.   This
may indicate that Moyal-Weyl deformations considered in the framework of 
twisted Poincar\'e covariance are too trivial for this scope.
%, but might be also an example of a more general phenomenon.

%It should entail an appropriate formulation of the twisted
%version R1$_\star$ of the  Lorentz covariance axiom R1; the latter should
%fix the free field $\star$-commutation relations. Although we have not
%presented a proposal for R1$_\star$, we have argued that it should imply W4
%(Lorentz covariance of Wightman function).
%It is possible to impose the usual locality axiom R2$_\star$. The consequences
%for the Wightman functions, in particular the cluster property W6,
%are the same as in the undeformed QFT.
%The $\star$-commutator of free fields coincides with the undeformed
%counterpart.

As a general remark,  we would like to emphasize that the cluster property W6 is
an important test for QFT on noncommutative as well as  on commutative spaces:
its violation implies a macroscopic (and therefore contrasting with
experiments) violation
of causality. It is also an easy theoretical test to carry out on free fields.
For the noncommutative space at hand, our two possible prescriptions for free fields fulfill
the cluster property whereas other prescriptions proposed
in the literature  ($\theta'\neq \theta$, see
the end of section \ref{Free}) lead to its violation.

As already noted in the paper, our results have some overlap 
or links with those of other works. To this regard we add
some further remarks.
At the 21st Nishinomiya-Yukawa Memorial Symposium
on Theoretical Physics (11-15 Nov. 2006) ``Noncommutative Geometry
and Quantum Spacetime in Physics'', after presenting our results,
the author of  Ref. \cite{Abe06} pointed out his work. We have realized that
there he proposes a quantization procedure  for scalar fields and 
$\theta^{0i}=0$
which finally coincides with the first of our two proposals and arrives
at very similar conclusions, although the derivation is different and
various steps of it appear to us not completely clear or justified.
In \cite{MadSte99} field quantization on the $h$-deformed Lobachevsky
plane was performed adopting a braided tensor product
among coordinates of different spacetime points, as done here; 
by a proper treatment the authors found that
the result for the 2-point function also coincides with the undeformed one.
Finally, already in \cite{Oec99,Oec00} Oeckl used the relation between 
the deformed and undeformed covariance to determine a mapping between 
deformed and undeformed theories
(in the Euclidean formulation of QFT); in the Moyal-Weyl 
case this mapping allows \cite{Oec00} to immediately compute
the deformed Green functions in terms of the undeformed ones 
 (however they do not coincide).

\subsection*{Acknowledgments}

G. F. acknowledges financial support by the Alexander von
Humboldt Stiftung for a stay at the ``Arnold Sommerfeld Center for
Theoretical Physics'' of the University LMU of Munich, where
the main part of this work has been done.
We are grateful to P. Aschieri, H. Steinacker, S. Doplicher, 
G. Piacitelli, F. Lizzi, P.
Vitale for useful and stimulating discussions.

\end{document}